\documentclass[aps,pra,twocolumn,showpacs,amsmath,floatfix,superscriptaddress]{revtex4}
\usepackage{epsfig}

\newcommand{\ohalf}{\mbox{$_\frac{1}{2}$}}
\newcommand{\thalf}{\mbox{$_\frac{3}{2}$}}
\newcommand{\fhalf}{\mbox{$_\frac{5}{2}$}}
\newcommand{\oh}{\mbox{$_{1/2}$}}
\newcommand{\thh}{\mbox{$_{3/2}$}}
\newcommand{\fh}{\mbox{$_{5/2}$}}
\newcommand{\ts}{\mbox{$\mbox{s}$}}
\newcommand{\tp}{\mbox{$\mbox{p}$}}
\newcommand{\td}{\mbox{$\mbox{d}$}}

\begin{document}
\title{
Interchannel coupling effects in the spin polarization
of energetic photoelectrons
}
\date{\today}
\author{Himadri S. Chakraborty}
\affiliation{
Max-Planck-Institut f\"ur Physik Komplexer
Systeme, N\"{o}thnitzer Strasse 38,
D-01187 Dresden, Germany
}
\affiliation{
James R. Macdonald Laboratory, Department of Physics, 
Kansas State University, Manhattan, Kansas 66506-2604
}
\author{Pranawa C. Deshmukh}
\affiliation{
Department of Physics, Indian Institute of Technology-Madras,
Chennai 600036, India
}
\author{Steven T. Manson}
\affiliation{
Department of Physics and Astronomy, Georgia State University,
Atlanta, Georgia 30303-3083
}

\begin{abstract}
Effects of the interchannel coupling on the spin polarization
of energetic photoelectrons emitted
from atomic Ne valence subshells are examined. Like  
previously obtained results for cross sections and angular distributions,
the photoelectron spin polarization parameters too are   
found considerably influenced by the coupling.  
The result completes a series of studies to 
finally conclude that the independent
particle description is inadequate 
for the {\em entire} range of photoionization
dynamics over the {\em full} spectral energy domain. 
\end{abstract}

\preprint{HE-PI/SpinP}
\maketitle

\section{\label{sec:intro}INTRODUCTION}

The study of the spin polarization of photoelectrons is 
important primarily on two counts. Firstly, since this effect 
originates from purely 
relativistic interactions, its behavior provides 
insights into 
relativistic aspects of the dynamical correlation which
are inaccessible by conventional studies of 
cross section and angular distribution; in the photoionization
of lighter atoms, which may normally seem tractable by a non-relativistic
approach, sizable resonance features in spin polarization
spectra of emerging electrons can be found when relativistic forces 
are included\cite{dias99,rado90}. Secondly, 
spin resolved spectroscopic measurements in conjunction with
cross section and angular distribution data provide
a comprehensive methodology to {\em completely} characterize the 
photoionization process\cite{kaem9193,snell01}. 
As a virtual beginning of the interest in the field, 
the emission of highly spin
polarized electrons over a limited range of the ejection angle
near the Cooper minimum of the photoionization cross section was 
predicted by Fano many years ago\cite{fano69}. Since then,
the spin polarization of photoelectrons emanating from
unpolarized atoms has been the subject of several theoretical
and experimental investigations (for reviews see
Refs.\,\cite{cher83,heinz96}).  
However, the focus of all these studies has been the low
photon energy range (VUV and soft X-ray), 
over which it is a common knowledge that
the electron correlation, in its complete form including
interchannel coupling, is significant and often dominating.

On the other hand, at photon energies far away from the ionization              
threshold, it was believed until recently that 
the independent particle (IP)
framework, which completely disregards electron correlation,
can adequately describe the photoionization 
process\cite{beth58,coop75,mans78,berk79,star82,star96}. 
However, over a series of combined experimental and theoretical  
studies this notion has recently been corrected for
the cross section and the angular distribution 
asymmetry parameter\cite{dias97,hans99,chak01}. 
Interchannel coupling 
has been shown to be a crucial determinant of the quantitative accuracy
of these parameters in the intermediate and the high energy 
regime for photoelectrons emitted from both 
inner and outer atomic subshells. From a perturbative
perspective, 
the effect originates from the correction to the single
channel matrix element from a continuum configuration interaction 
among all neighboring 
channels\cite{dias97}. 
Since the dynamical difference among relativistic 
(spin-orbit) channels arising from a given subshell determines the
spin polarization character of electrons photoejected
from that subshell,
it is of particular interest to examine
how the spin polarization parameters are affected
by correlation in the form of interchannel coupling mechanism.
In this paper, we focus on this aspect  
by investigating the spin polarization parameters of 
the valence 2p photoelectrons 
from atomic Ne.   

\section{\label{sec:math}ESSENTIAL THEORETICAL DETAILS}

The relativistic-random phase approximation (RRPA)\cite{john79,john80}
has been employed to perform the calculation. The RRPA calculation 
starts from an explicitly relativistic basis so that relativistic 
interactions are included {\em ab\,initio}.  In
addition to the ground state correlation, as well as two-electron
promotion in the residual Ne-ion core, RRPA incorporates interchannel
coupling among {\em all} of the single excitation/ionization 
final state channels.
We use a
framework in which the coupling among selective members 
of the relativistic dipole-allowed
$jj$-coupled channels:
\begin{eqnarray}\label{channel}
\begin{array}{lllrlll}
1\ts\ohalf & \rightarrow & k\tp\thalf, k\tp\ohalf,&~~~~~
&2\ts\ohalf & \rightarrow & k\tp\thalf, k\tp\ohalf,  \\
2\tp\ohalf & \rightarrow & k\td\thalf, k\ts\ohalf,&\mbox{and~~~} 
& 2\tp\thalf & \rightarrow & k\td\fhalf, k\td\thalf,
k\ts\ohalf
\end{array}                                                                    
\end{eqnarray}
can be chosen.  
The calculation has been carried out
in both the length and the velocity gauge formalism; the good agreement
between
length and velocity results, even at highest photon 
energy considered,
indicates the numerical accuracy of our calculation. It may be 
mentioned here that RRPA has been used previously with
reasonable success to study the spin polarization of
photoelectrons from noble gases, but only in the low 
photon energy range\cite{huan79}.  

In this calculation, we have considered the case 
where the target atom is unpolarized and
the polarization of the residual ion is not observed. Equivalently,
the polarization of the target atom is averaged out and that of the 
residual ion is summed over.
The dipole photoionization can then be {\em completely} described, in
general,
by a set of five dynamical parameters $\sigma$, $\beta$, $\xi$, $\eta$,
and
$\zeta$, wherein $\sigma$ is the partial cross section,
$\beta$ is the angular distribution asymmetry parameter and the others
are the photoelectron spin polarization parameters. These dynamical
parameters can be expressed in terms of the reduced dipole matrix
elements\cite{john79} of the process. 
The explicit expressions for the spin
polarization parameters corresponding to both $2\tp\ohalf$ and
the $2\tp\thalf$ photoelectrons of Ne are given as \cite{huan80}
\begin{widetext}
\begin{subequations}
\FL
\begin{eqnarray}\label{xi3h}
\xi_{2p\thalf}&=&\left[
\frac{1}{2}\left|D_{\ts\oh}\right|^2+\frac{2}{5}\left|D_{\td\thh}\right|^2
 -\frac{9}{10}\left|D_{\td\fh}\right|^2
 -\frac{\sqrt{5}}{2}\left|D_{\ts\oh}\right|\left|D_{\td\thh}\right|
  \cos\left(\theta_{\ts\oh}-\theta_{\td\thh}\right)
                 \right.\nonumber\\
                 &&\left.
 +\frac{9}{10}\left|D_{\td\thh}\right|\left|D_{\td\fh}\right|                  
  \cos\left(\theta_{\td\thh}-\theta_{\td\fh}\right)
                  \right]\times \overline{\sigma}_{2\tp\thalf}^{-1}
\end{eqnarray}
\FL
\begin{equation}\label{xi1h}
\xi_{2p\ohalf}=\left[
 -\left|D_{\ts\oh}\right|^2+\left|D_{\td\thh}\right|^2
 -\frac{1}{\sqrt{2}}\left|D_{\ts\oh}\right|\left|D_{\td\thh}\right|
  \cos\left(\theta_{\ts\oh}-\theta_{\td\thh}\right)
                  \right]\times \overline{\sigma}_{2\tp\ohalf}^{-1}
\end{equation}
\end{subequations}
\begin{subequations}
\FL
\begin{eqnarray}\label{eta3h}
\eta_{2p\thalf}&=&\left[
-\frac{3}{2}\sqrt{\frac{1}{5}}\left|D_{\ts\oh}\right|\left|D_{\td\thh}\right|
   \sin\left(\theta_{\ts\oh}-\theta_{\td\thh}\right)
  +3\sqrt{\frac{1}{5}}\left|D_{\ts\oh}\right|\left|D_{\td\fh}\right|
   \sin\left(\theta_{\ts\oh}-\theta_{\td\fh}\right)
                    \right.\nonumber\\
                    &&\left.
  -\frac{3}{2}\left|D_{\td\thh}\right|\left|D_{\td\fh}\right|
   \sin\left(\theta_{\td\thh}-\theta_{\td\fh}\right)
                    \right]\times \overline{\sigma}_{2\tp\thalf}^{-1}
\end{eqnarray}
\FL
\begin{equation}\label{eta1h}
\eta_{2p\ohalf}=\left[
   \frac{3}{\sqrt{2}}\left|D_{\ts\oh}\right|\left|D_{\td\thh}\right|
   \sin\left(\theta_{\ts\oh}-\theta_{\td\thh}\right)
                  \right]\times \overline{\sigma}_{2\tp\ohalf}^{-1}
\end{equation}
\end{subequations}
\begin{subequations}
\FL
\begin{eqnarray}\label{zeta3h}
\zeta_{2p\thalf}&=&\left[
-\frac{1}{2}\left|D_{\ts\oh}\right|^2+\frac{1}{5}\left|D_{\td\thh}\right|^2
 +\frac{3}{10}\left|D_{\td\fh}\right|^2
 -\sqrt{5}\left|D_{\ts\oh}\right|\left|D_{\td\thh}\right|
  \cos\left(\theta_{\ts\oh}-\theta_{\td\thh}\right)
                 \right.\nonumber\\
                 &&\left.
 -\frac{9}{5}\left|D_{\td\thh}\right|\left|D_{\td\fh}\right|
  \cos\left(\theta_{\td\thh}-\theta_{\td\fh}\right)
                  \right]\times \overline{\sigma}_{2\tp\thalf}^{-1}
\end{eqnarray}
\FL
\begin{equation}\label{zeta1h}
\zeta_{2p\ohalf}=\left[
  \left|D_{\ts\oh}\right|^2+\frac{1}{2}\left|D_{\td\thh}\right|^2
 -\sqrt{2}\left|D_{\ts\oh}\right|\left|D_{\td\thh}\right|
  \cos\left(\theta_{\ts\oh}-\theta_{\td\thh}\right)
                  \right]\times \overline{\sigma}_{2\tp\ohalf}^{-1}
\end{equation}
\end{subequations}
where, with the photon energy $\omega$, the subshell cross sections
are
\begin{subequations}
\begin{eqnarray}\label{sig3h}
\sigma_{2p\thalf}=\frac{8\pi^4}{\omega c}\overline{\sigma}_{2\tp\thalf}
         &=& \frac{8\pi^4}{\omega c}\left[\left|D_{\ts\oh}\right|^2
         + \left|D_{\td\thh}\right|^2 + \left|D_{\td\fh}\right|^2\right]
\end{eqnarray}
\begin{eqnarray}\label{sig1h}
\sigma_{2p\ohalf}=\frac{8\pi^4}{\omega c}\overline{\sigma}_{2\tp\ohalf}
         &=& \frac{8\pi^4}{\omega c}\left[\left|D_{\ts\oh}\right|^2
         + \left|D_{\td\thh}\right|^2\right]
\end{eqnarray}
\end{subequations}
\end{widetext}
In the above equations we use the short-hand notations 
$D_{l'_{j'}}$ and $\theta_{l'_{j'}}$ for the reduced matrix 
element $D_{nl_j\rightarrow kl'_{j'}}$ and the phase-shift 
$\theta_{nl_j\rightarrow kl'_{j'}}$ respectively,
corresponding to the $nl_j\rightarrow kl'_{j'}$ dissociation channel.
Conventionally, additional parameters 
$\delta_{nl_j}=(\zeta_{nl_j}-2\xi_{nl_j})/3$ 
are also used which connect to the spin polarization of the total 
photoelectron flux\cite{huan80}. 

\section{\label{resdis}RESULTS AND DISCUSSION}

In order to uncover the details of how the interchannel coupling
influences the 
photoelectron spin polarization dynamics we have performed five separate
calculations for
the 2p$\thalf$ and 2p$\ohalf$ photoionization of Ne with varying degrees 
of interchannel coupling. These are (i) no interchannel coupling
among channels arising from different relativistic subshells, (ii)
coupling of all channels 
from 2p$\thalf$ and $2p\ohalf$ subshells, (iii) from
2p and 2s subshells, (iv) from 2p and 1s subshells, and (v)from
all four subshells together (full calculation). Since, 
in calculation (i), we ignore all
effects of interchannel coupling between channels arising from differing 
relativistic subshells, this is similar to the relativistic independent
particle (IP) description, except that ground state correlations are
included 
along with coupling among channels arising from the same subshell.
   
An elegant approach to understand the relative importance of
the coupling with different neighboring channels 
has been described in Ref.\,\cite{dias97} in the spirit of
a first order perturbation theory. 
Under the influence of a perturbing degenerate channel $J$
the corrected wavefunctions $\Psi_{{2p}_j}$ for any dipole
channel $2\tp_j\rightarrow k\ts_{1/2}(k\td_{j'})$ 
(see Eq.\,(\ref{channel})) from either of  
$2\tp_j (j=1/2,3/2)$ subshells are given, 
at the photoelectron kinetic energy $E$, by 
\begin{widetext}
\begin{eqnarray}\label{ic_wave}
\Psi_{{2p}_j}(E) &=&
    \psi_{{2p}_j}(E)
    +\int \frac{\langle\psi_{J}(E')
     \left|H-H_0\right|\psi_{{2p}_j}(E)\rangle}
     {E-E'}\psi_{J}(E') dE', 
\end{eqnarray} 
where $\psi$'s denote unperturbed wavefunctions, which are 
eigenfunctions of the unperturbed Hamiltonian $H_0$, 
and the final state total
angular momentum $j'$ has two dipole allowed values 5/2 and 3/2.
In Eq.\,(\ref{ic_wave}), the matrix element under the energy integration
is the interchannel coupling matrix element with $H$ being the {\em full}
Hamiltonian of the system. 
Now, defining the dipole photoionization 
matrix element for $2\tp_j\rightarrow k\ts_{1/2}(k\td_{j'})$ transitions 
with no interchannel coupling among channels arising from different 
relativistic subshells, corresponding to calculation (i), as 
\begin{equation}\label{ip_amp}
D_{{2p}_j}(E)=\langle\psi_i|T|
     \psi_{{2p}_j}(E)\rangle,
\end{equation}
with $\psi_i$ being the ground state wavefunction and $T$ the transition
operator, the corresponding perturbed matrix elements  
can be expressed as
\begin{eqnarray}\label{ic_amp}
M_{{2p}_j}(E) &=&
    D_{{2p}_j}(E)
    +\int \frac{\langle\psi_{J}(E')
     \left|H-H_0\right|\psi_{{2p}_j}(E)\rangle}
     {E-E'}D_{J}(E') dE'.
\end{eqnarray} 
\end{widetext}
The correction term on the right side of Eq.\,(\ref{ic_amp}) can be
significant 
if two conditions are simultaneously satisfied. 
First, the spatial overlap between the perturbed and perturbing 
channel wavefunctions must
be considerable to result in a significant interchannel coupling matrix 
element; this is expected when the discrete wavefunctions have the same 
principle quantum numbers so that they occupy the same region of space 
and have significant overlap, and the respective ionization thresholds are
close
so that at high enough energies the electrons from both subshells
have similar momenta, which enable the continuum wavefunctions to
oscillate
roughly ``in phase''.
Second, the magnitude of the unperturbed matrix element of the perturbing
channel
is considerably larger than that of the perturbed channel. 
\begin{figure}
\centerline{\psfig{figure=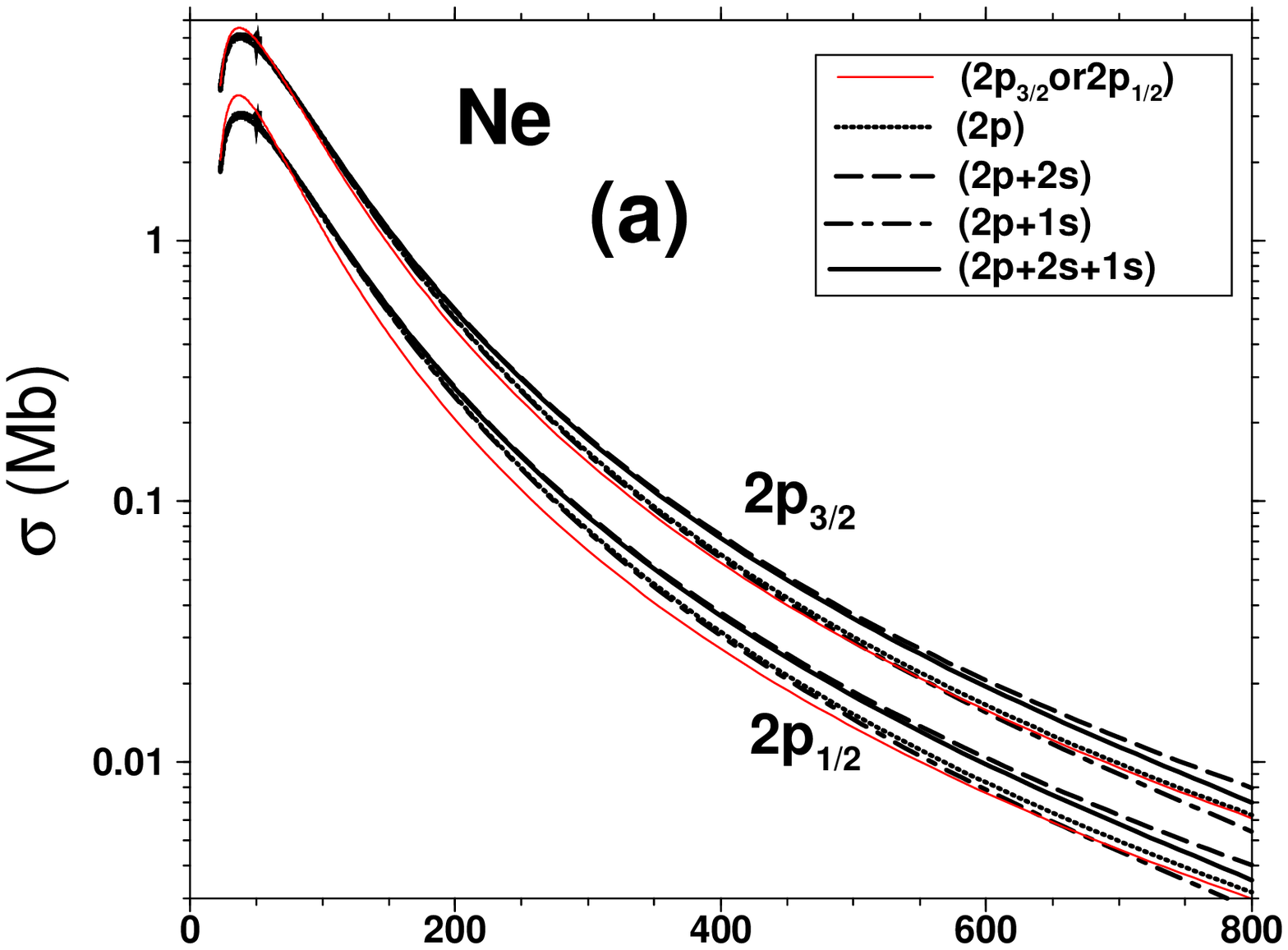,height=4.7cm,width=9.0cm,angle=0}}
\centerline{\psfig{figure=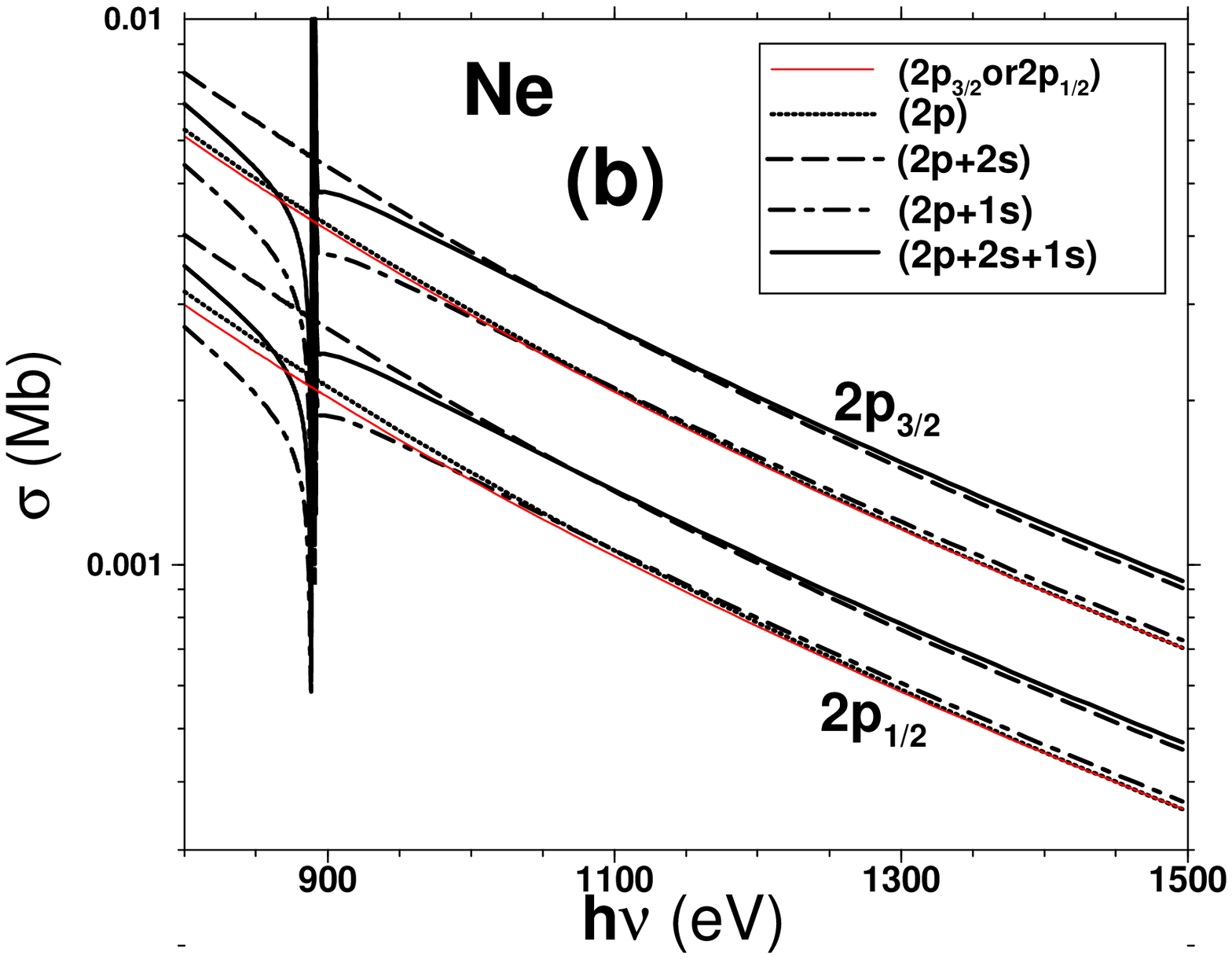,height=5.2cm,width=9.0cm,angle=0}}
\caption{\label{sigma}Ne 2p\thalf and 2p\ohalf spin-orbit subshell 
cross sections in
several
selections of channels calculated by the relativistic-random phase
approximation: (a) for photon energy up to 800 eV and (b) from 800 eV
to 1.5 KeV. The structure around 900 eV is due to 1s Rydberg
resonances.}
\end{figure}
Now the form of the energy integral suggests that the primary contribution
of interchannel 
interaction will come 
from the values of the integrand at $E'\simeq E$. Importantly further, 
the electron continuum wavefunctions, participating
in the energy integral, must be normalized per unit energy
through a multiplication by a factor $m\hbar^{-2}E^{-1/4}$ \cite{fano86}. 
Therefore, the leading energy 
behavior of the interchannel coupling matrix
element in Eq.\,(\ref{ic_wave}) turns out to be $E^{-1/2}$
when the energy is high enough.
Evidently, considering Eq.\,(\ref{ic_amp}),  
if the uncoupled matrix element $D_{J}$ of channel $J$
decreases with energy slower by a factor $E^{1/2}$ or more than the
corresponding
decay of $D_{{2p}_j\rightarrow k\ts_{1/2}(k\td_{j'})}$, the resulting
effect of the coupling will be considerable, provided 
the interchannel coupling matrix element is significant.  
Indeed, this leads us to expect a 
strong effect of the 2s channels (with Dirac-Fock threshold 52.68 eV)
on 2p$_j$ photoionization (with thresholds 23.08 eV for $j=3/2$ 
and 23.21 eV for $j=1/2$).  This is clearly seen in Fig.\,1 which gives 
our calculated results for the $2p\oh$ and the $2p\thh$ subshell cross
sections 
in each of the five calculations described above; at the highest energies, 
the cross section results are seen to essentially coalesce into just two
curves - 
those including coupling between 
and those 
omitting that coupling form the other.  This was noted previously
in Ref.\,\cite{dias97} 
where similar calculations were performed.
Because the high energy uncoupled photoionization cross section
for an nl subshell falls off with energy as $E^{-(7/2+l)}$, 
the correction term in Eqs.\,(\ref{ic_amp}) falls off in
the limit of
$\omega \rightarrow E$, as $E^{-(5/4+l/2)}$)\cite{comment}.  As applied to
the 
present case, the perturbation of $D_{{2p}_j}$ by $D_{2s}$ falls off as
$E^{-(5/4+l/2)}$,
the same as the falloff of the $D_{{2p}_j}$ themselves, so that the
perturbation is of 
the same order of size as the uncoupled matrix elements.
The weak effect
of 1s channels on either of 2p$_j$ cross sections, as also seen in
Fig.\,1, 
is owing to the much higher   
1s ionization threshold (893.02 eV) that results in   
poor overlap of the continuum wavefunctions in 
ensuing interchannel coupling matrix element, along with the fact that the
discrete 
1s and 2p$_j$ occupy very different regions of space and, therefore,
overlap poorly. 

In addition, what is rather interesting to note in Fig.\,1 is the tiny
effect from
the coupling between the channels from the spin-orbit split 2p$_j$
subshells. The curve (thin solid) corresponding to only-2p$\thalf$ or
only-2p$\ohalf$  
channels (calculation (1), similar to the IP result) differs very little
in the lower part
of the energy range (Fig.\,1(a)) when compared to the curve (dotted)
from all 2p channels combined; however, both the curves practically merge
together at higher energies (Fig.\,1(b)) indicating virtually no effect
from the coupling. This phenomenon can be 
understood as follows. It is true that the interchannel coupling 
matrix element between the 2p$\thalf$ and 2p$\ohalf$ channels is 
strong due to the close proximity of their respective ionization
thresholds. But since the high energy falloff of 
$D_{{2p}_{3/2}}$ and $D_{{2p}_{1/2}}$ are the same, the ensuing
coupling corrections (Eq.\,(\ref{ic_amp})) fall off effectively 
as $E^{-1/2}$ at higher energy, explaining how the small coupling effect
at
lower energies becomes practically
zero at higher energies. 

Looking at our results for the spin polarization parameters,
Figs.\,2 through 5, rather different phenomenology is evident; significant
effects 
resulting from the coupling 
between the channels arising from the 2p$\thh$ and 2p$\oh$ spin-orbit
subshells are noted!
Understanding the underlying reason(s) for this phenomenology is somewhat 
more complex than for the cross sections owing to the fact that the 
spin polarization parameters depend upon both the magnitudes {\em and} the 
phases of the dipole matrix elements, as seen from Eqs.\,(2-4).  Thus, 
an understanding of the modification of the phase shifts engendered by
interchannel 
coupling is also of importance.
 
Starting from Eq.\,(\ref{ic_amp}) and
explicitly introducing the unperturbed ($\theta$) and
the perturbed ($\Theta$) phase-shift through 
the notations $D =
|D|\exp(i\theta)$ and $M =
|M|\exp(i\Theta)$ we obtain 
\begin{widetext}
\begin{eqnarray}\label{ic_phase}
\Theta_{2\tp_j}\! &=&\! \theta_{2\tp_j}
\!-i\log\left[\!\left(\left|D_{{2p}_j}\right|\!(E)
    +\!\int\!\frac{\langle\psi_{J}(E')
     \left|H-H_0\right|\psi_{{2p}_j}(E)\rangle}
 {E-E'}\left|D_{J}\right|\exp[i(\theta_J-\theta_{2\tp_j})
 ](E')dE'\right)
 \times\left|M_{{2p}_j}\right|^{-1}\right].
\end{eqnarray}
\end{widetext}
Thus, as discussed above, sufficiently above the ionization 
thresholds the leading energy
behavior of $\langle{\psi}_J\left|H-H_0\right|
{\psi}_{2\tp_j\rightarrow J_\pm}\rangle$
is $E^{-1/2}$. From Eq.\,(\ref{ic_phase}) this indicates the decreasing
effect of the coupling on the relative phase-shift going up in the energy. 
However, it is important
to note here that this decay is much slower than $E^{-1/2}$ due to 
the logarithmic nature of the correction ---
a behavior which bears some consequence in the interchannel 
coupling effects on the photoelectron
spin polarization parameters.
\begin{figure}
\centerline{\psfig{figure=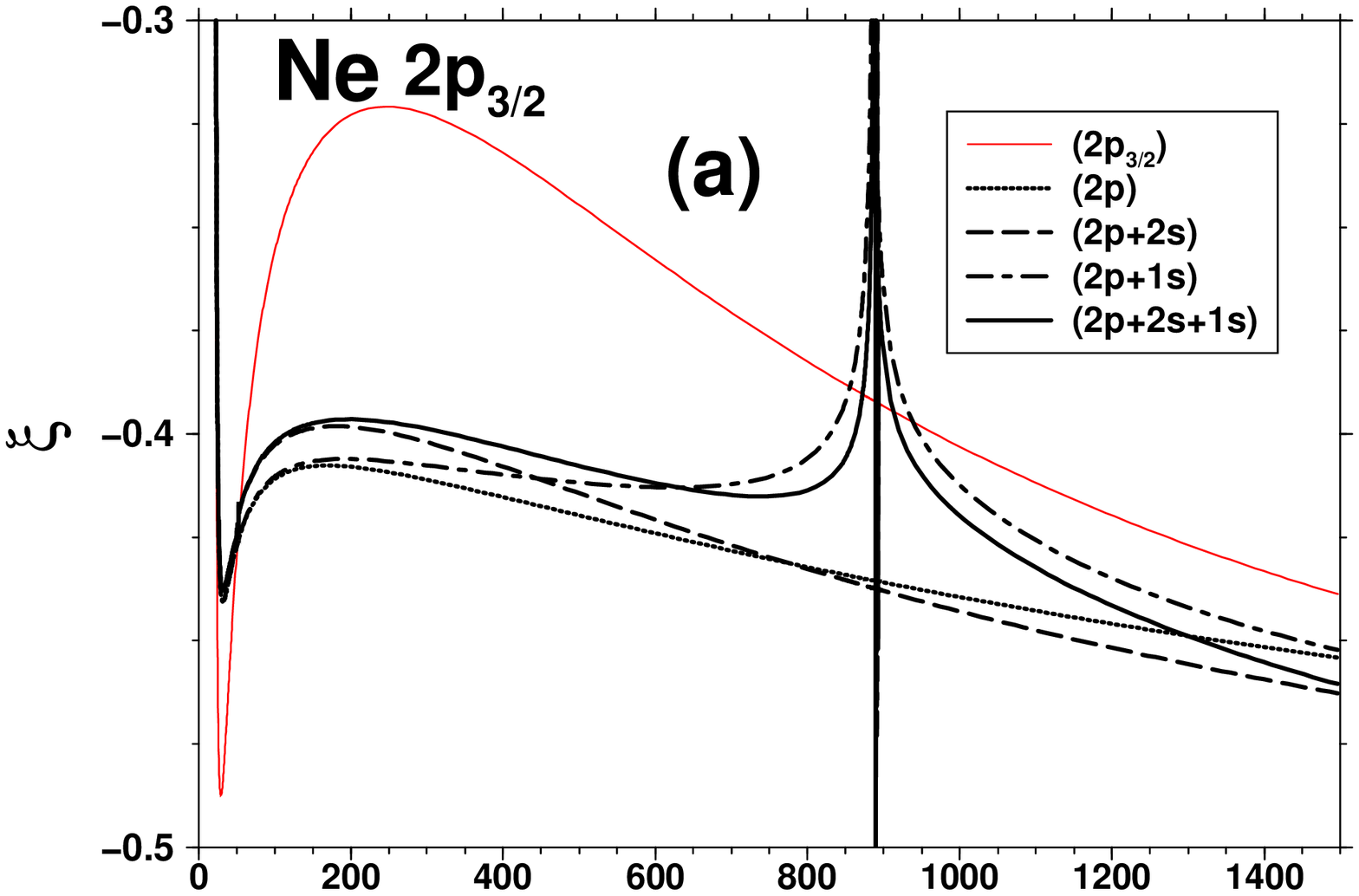,height=4.7cm,width=8.5cm,angle=0}}
\centerline{\psfig{figure=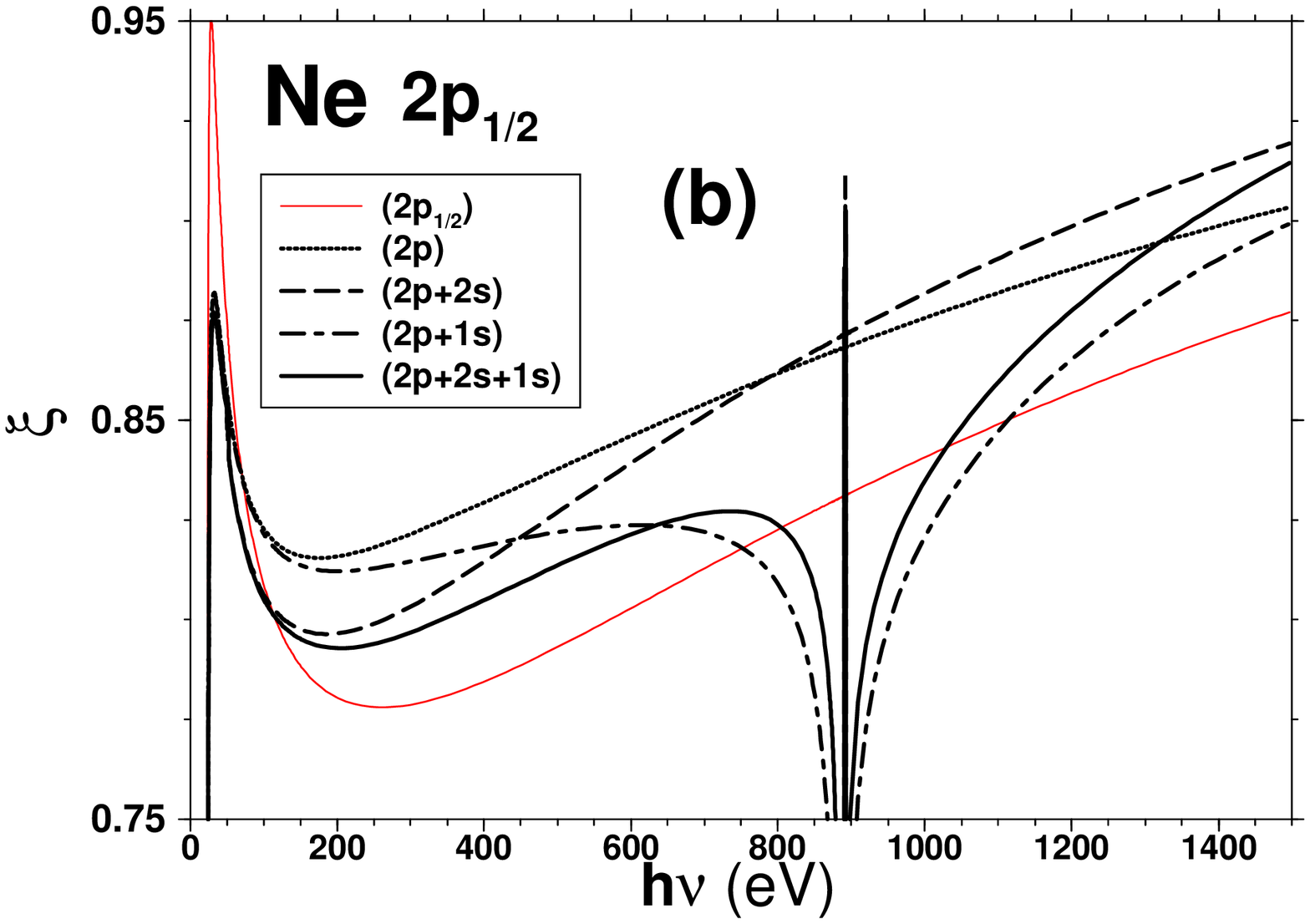,height=5.2cm,width=8.5cm,angle=0}}
\caption{\label{xi}Spin polarization parameter $\xi$ for (a) 2p\thalf and (b)
2p\ohalf
photoelectrons calculated in the same selections of channels as in Fig.\,1.}
\end{figure}
\begin{figure}
\centerline{\psfig{figure=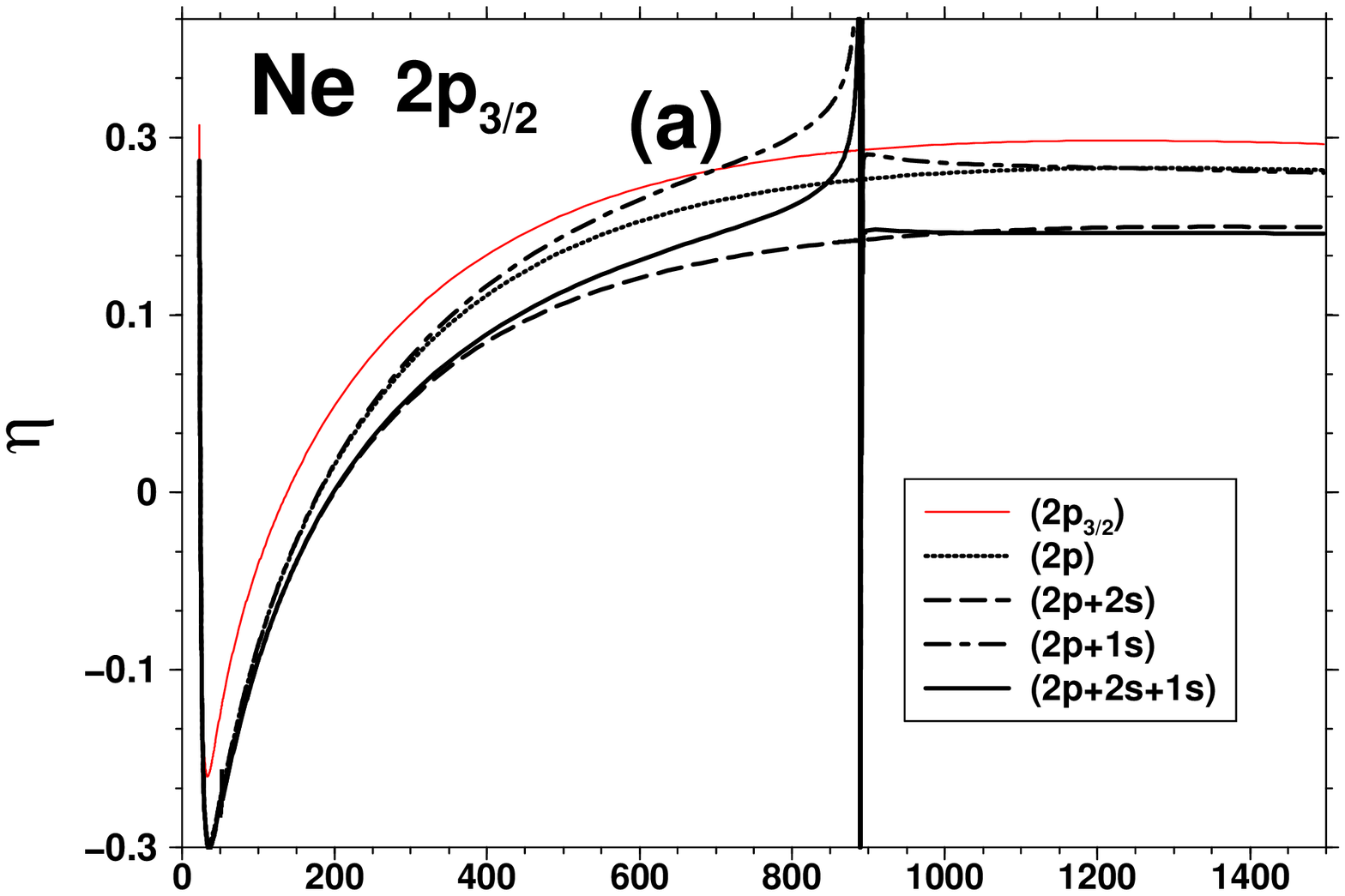,height=4.7cm,width=8.5cm,angle=0}}
\centerline{\psfig{figure=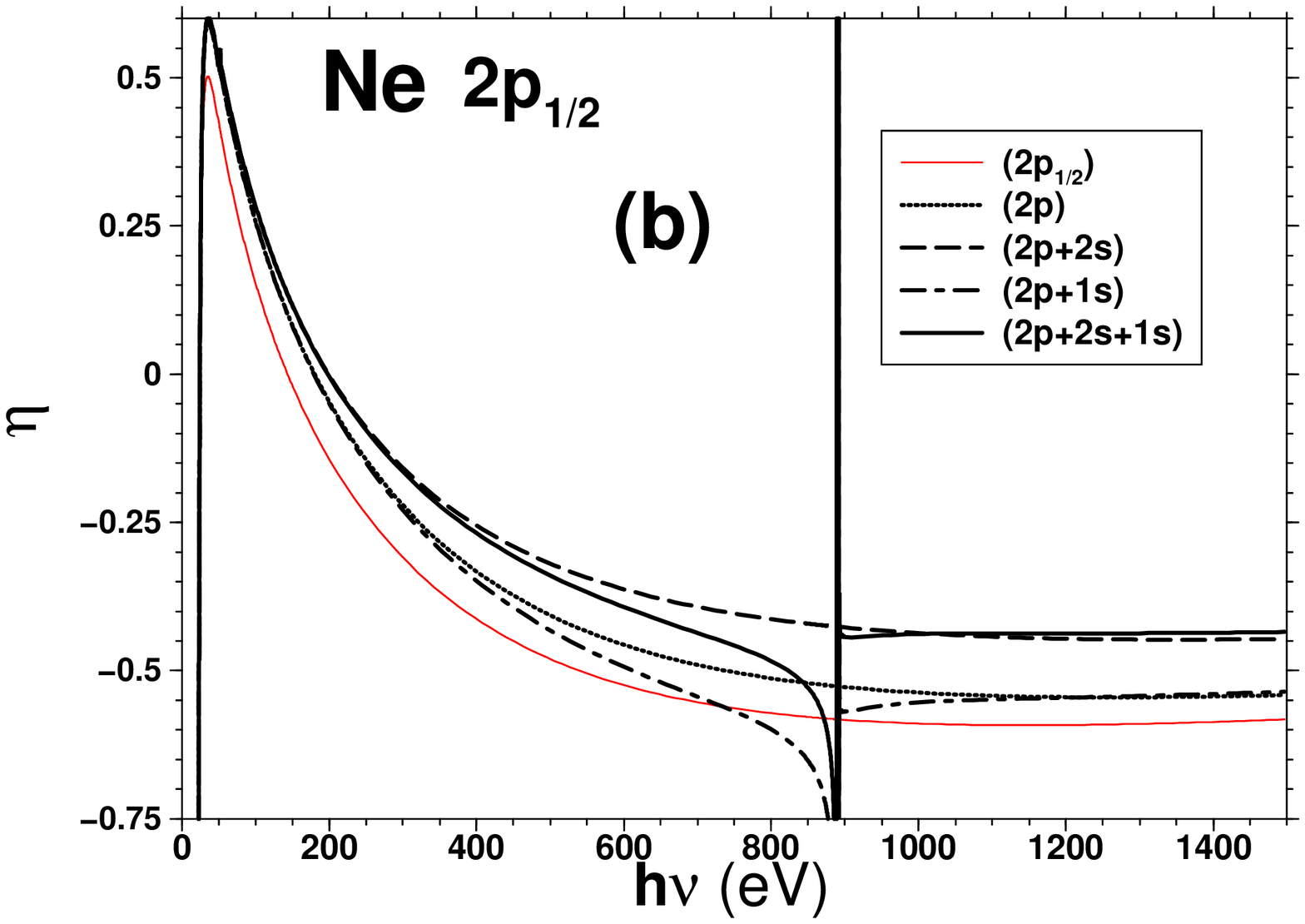,height=5.2cm,width=8.5cm,angle=0}}
\caption{\label{eta}Spin polarization parameter $\eta$ for (a) 2p\thalf and (b)
2p\ohalf
photoelectrons.}
\end{figure}

Let us first focus on the effect of 2p$\ohalf$  channels on the spin
polarization 
of 2p$\thalf$  photoelectrons and {\em vice\,versa} (dotted curves). We
compare between
the thin solid curve (effectively the IP prediction) and the dotted curve
of each of
the Figs.\,2 through 5. Evidently, for all the spin polarization
parameters
the effect of this coupling is stronger at relatively low energies.  This
is clearly 
because the strength of the interchannel coupling matrix element decreases
with 
increasing energy, as discussed above.  But, as already made clear from
the corresponding 
cross section results, this coupling does not strongly affect the
magnitude of the 
2p$_j$ dipole matrix elements.  And since their unperturbed phases 
are nearly equal,
the coupling 
does not alter their phases much (see Eq.\,(\ref{ic_phase})).  
As a consequence, our results (not shown) of the angular distribution
asymmetry parameter $\beta$, that depends on both the matrix elements and
the phases, shows minimal effect of this coupling.
But if the coupling affects neither the
dipole matrix 
elements nor the phases significantly, 
how then can it affect the spin polarization
parameters?  The 
answer lies in the fact that the values of the 2p$_j$ spin polarization
parameters 
arise from complicated combinations of the dipole matrix elements and
their phases, 
Eqs.\,(2-4), resulting in significant cancellations, so that small
differences in 
dipole matrix elements and phases can be magnified to produce the results seen.
The fact that the non-relativistic limit of spin polarization parameters
are zero while for the $\beta$ parameter it is finite and close to
its relativistic value indicates that such a cancellation mechanism 
is indeed operative for the spin polarization parameters.  

For all of the spin polarization parameters, the result of 
this coupling, however, exhibits a rather slow
monotonic tendency to converge to the corresponding effective IP-like
prediction,
with increasing energy, much slower than the convergence of the cross
section.  
This is because the 
spin polarization parameters depend upon phase shift differences as well,
and these 
were shown above to converge more slowly than the magnitudes of the dipole
matrix 
elements.  This was also seen earlier in connection with $\beta$ which
also 
depends on phase shifts\cite{dias97}.  Further, note that 
this coupling induces in general a stronger influence on the spin
polarization of
2p$\thalf$ electrons than on that of 2p$\ohalf$ electrons, except for
$\eta$ (Fig.\,3)
where the results of both subshells have similar effect from this
coupling.
To provide some quantitative estimates, $\xi$ (Fig.\,2(a)) and 
$\delta$ (Fig.\,5(a)) for 2p$\thalf$ 
electrons show respectively about 30\% and 40\% modification compared to
the uncoupled results at roughly 300 eV photon energy; as expected, 
these differences 
decrease gradually as the energy increases. On the other hand,  
$\zeta$ for 2p\ohalf electrons (Fig.\,4(b), that is related to
the corresponding $\beta$ simply through $\zeta_{2\tp_{1/2}} = 
1- \beta_{2\tp_{1/2}}/2$, shows an almost negligible
effect
from this coupling.
\begin{figure}
\centerline{\psfig{figure=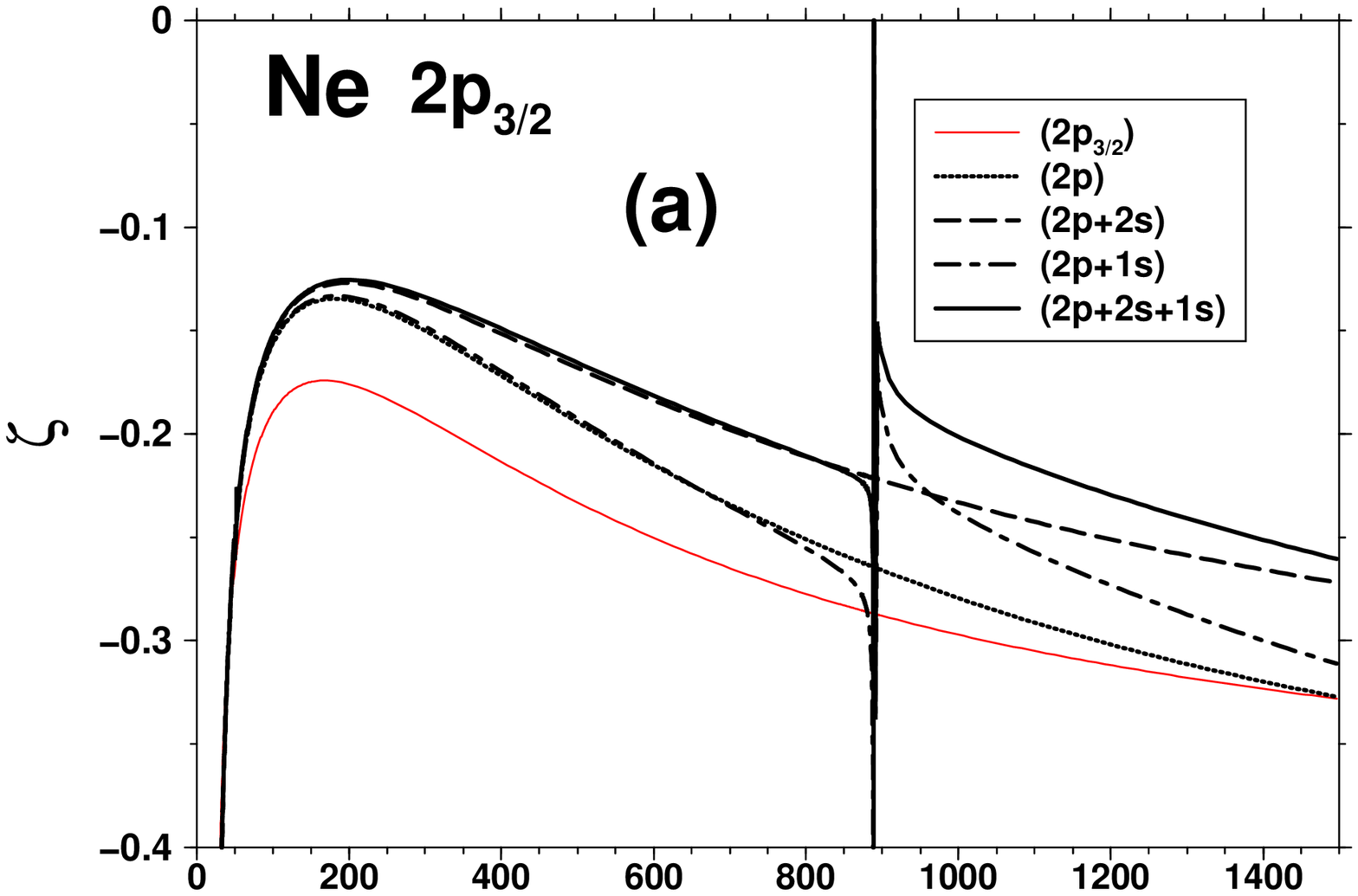,height=4.7cm,width=8.5cm,angle=0}}
\centerline{\psfig{figure=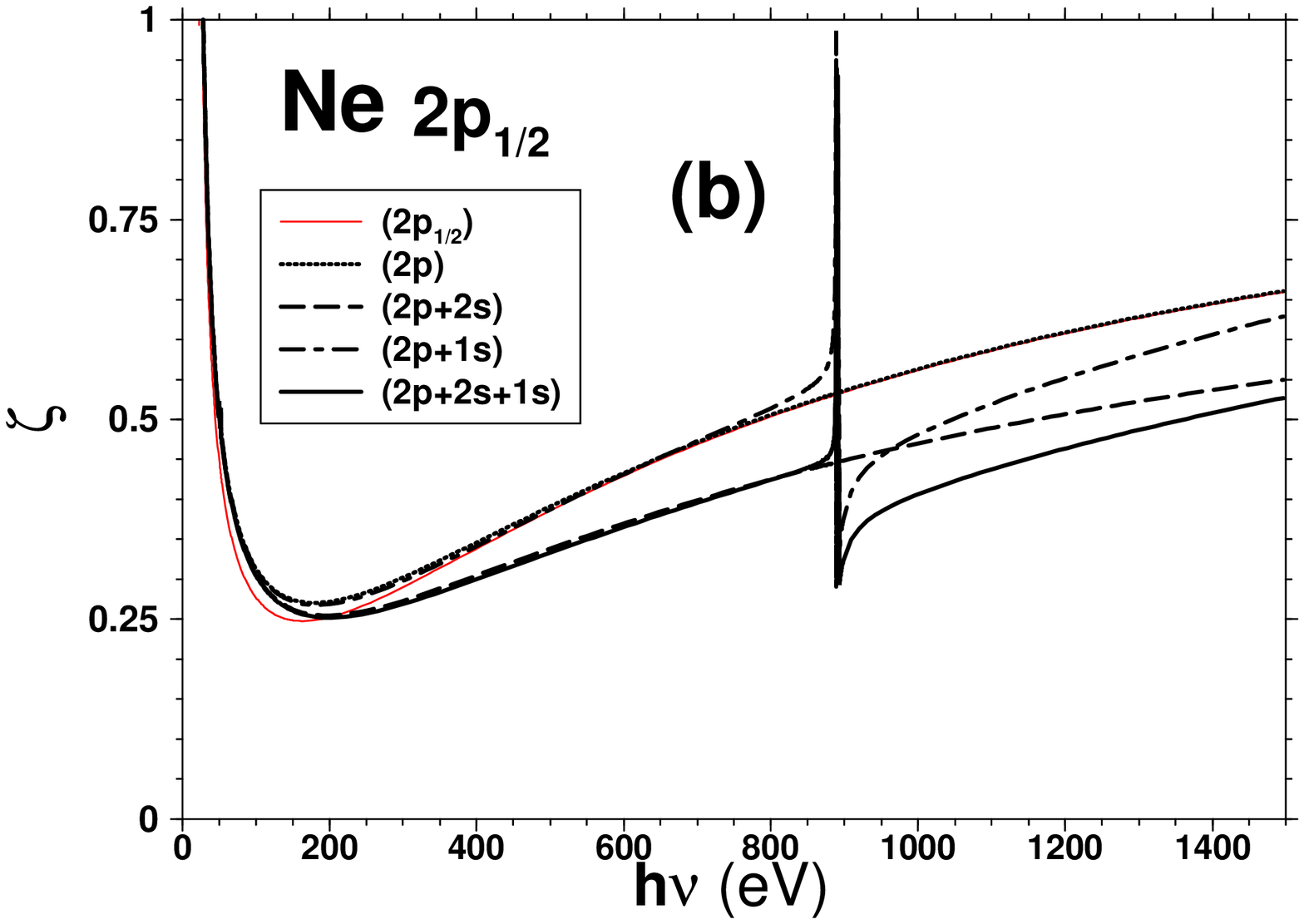,height=5.2cm,width=8.5cm,angle=0}}
\caption{\label{zeta}Spin polarization parameter $\zeta$ for (a) 2p\thalf and 
(b) 2p\ohalf
photoelectrons.}
\end{figure}
\begin{figure}
\centerline{\psfig{figure=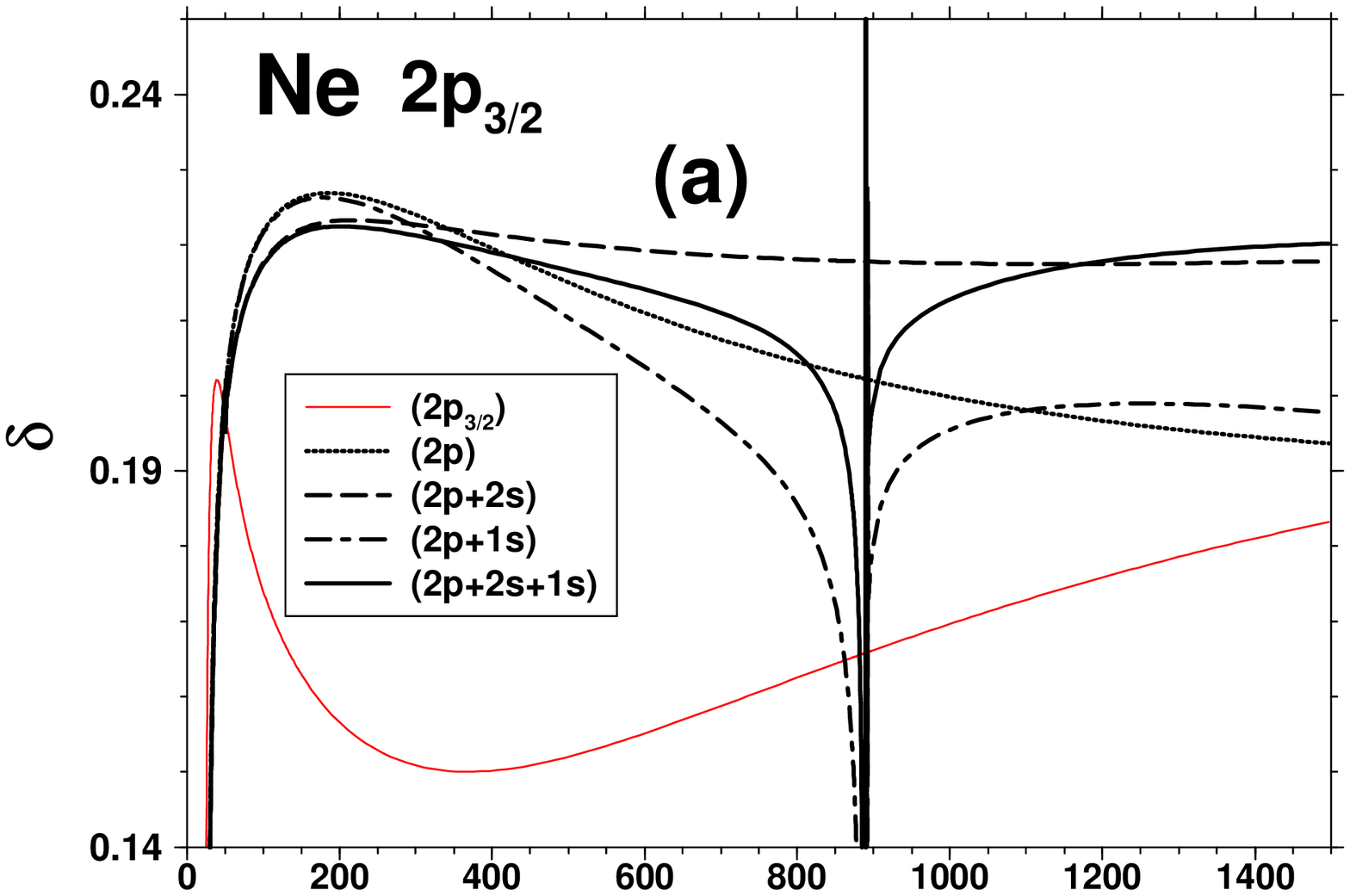,height=4.7cm,width=8.5cm,angle=0}}
\centerline{\psfig{figure=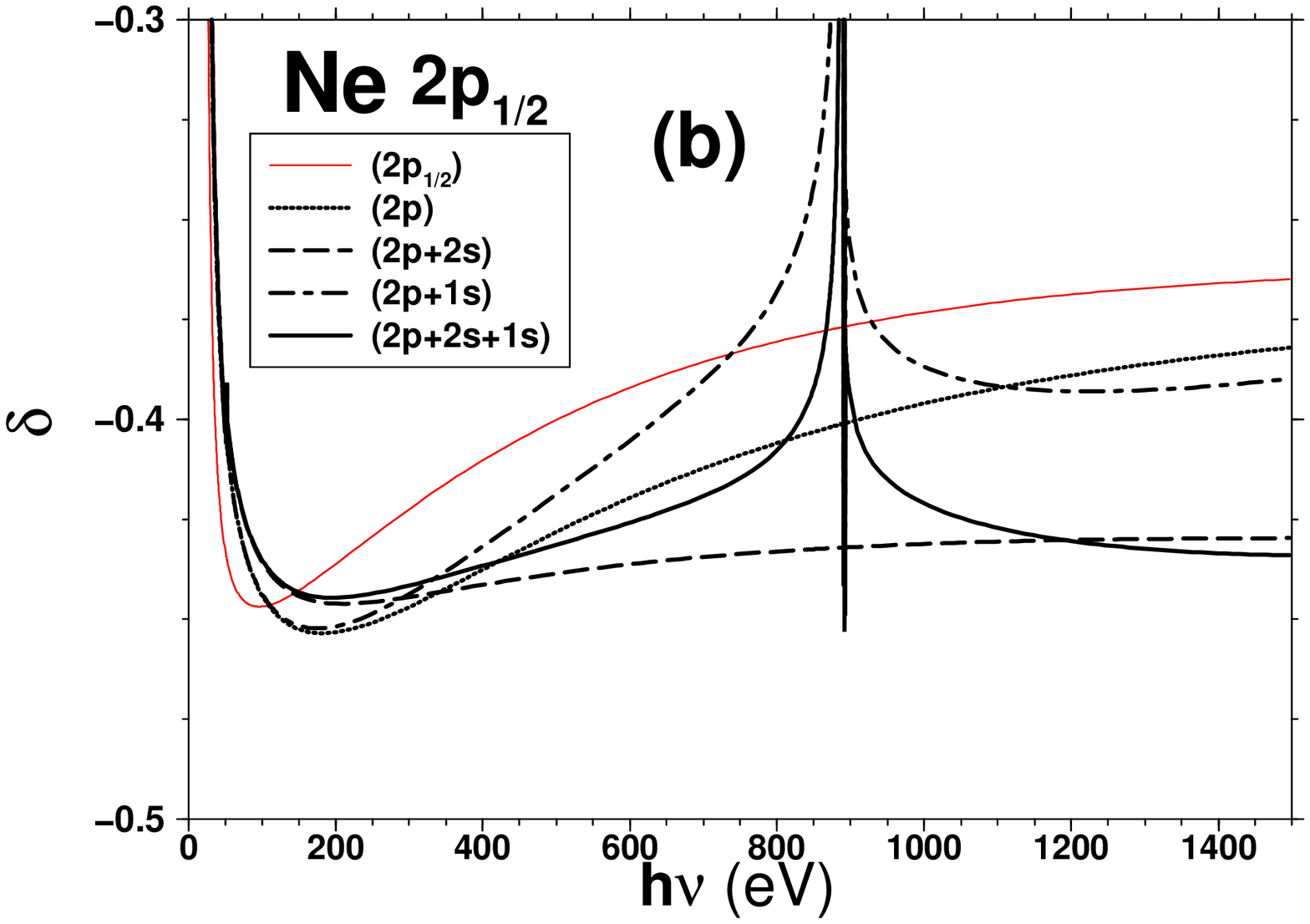,height=5.2cm,width=8.5cm,angle=0}}
\caption{\label{delta}Spin polarization parameter $\delta$ for (a) 2p\thalf and
(b) 2p\ohalf
photoelectrons.}
\end{figure}

Interchannel coupling of the 2p channels with either 2s or 1s channels 
involves alterations in both the magnitudes and phases of the 2p dipole 
matrix elements.  Coupling with the 2s channels affects the magnitudes 
of the 2p$\thalf$ and 2p$\ohalf$ dipole matrix elements very strongly, as
clearly indicated in Fig.\,1.  But the phases are also strongly affected
(see Eq.\,(\ref{ic_phase})).  
With the 1s coupling on the other hand, while the modifications to the
magnitudes of the 2p$_j$ dipole matrix elements are 
already small, except in a very small region around the 1s threshold,
the alterations of the 2p$_j$ phase shifts are also small.
But it is non-trivial to assess in which direction the 
changes in 
these dynamical quantities, the magnitudes and the phases of the 
2p$_j$ dipole matrix elements, will induce changes in the spin
polarization 
parameters.  For two of the spin polarization
parameters, $\xi$ (Fig.\,2) and $\delta$ (Fig.\,5), 
the effect of 2s and 1s coupling on 2p$_j$ ionization
is certainly not as straightforward
as in the case of cross sections where only the magnitudes of the
matrix 
elements (and not the phases) are important.
In addition, by virtue of the different
functional dependence of the parameters on the magnitudes of the dipole 
matrix elements and phase-shifts,
the qualitative behavior of the result changes from one parameter to
another.

As seen in Fig.\,2, the effect of 1s and 2s coupling on 2p$_j$ for the
parameter $\xi$
are roughly complementary until about 900 eV photon energy, the 
position of 1s Rydberg resonances; as a result, the deviation
between the dotted (all 2p channels included) and thick solid (full
calculation)
results remain approximately constant. Beyond the resonance region
the relative effect of
1s coupling drops off. However, over the entire energy range considered,
the full calculation differs from the effective IP result
(thin solid curve). This difference is significant for 2p$\thalf$ showing
a maximum
alteration of about 25\% at 300 eV, while for 2p$\ohalf$ the difference
is
rather small.

For parameters $\eta$ and $\zeta$, on the other hand, the 1s and 2s
coupling
influence the result quite in the similar qualitative manner as they do
for
the cross section. Along almost the complete energy range, albeit the
near-threshold region (where interchannel coupling is known to be 
important), 
$\eta$ for both 2p$\thalf$ and 2p$\ohalf$ (Figs.\,3(a,b)) as well as 
$2p\thalf$ $\zeta$ (Fig.\,4(a)) suggest an almost steady coupling
contribution of more than 25\% over the corresponding
effective IP prediction. For 2p$\ohalf$ $\zeta$ (Fig.\,4(b)), of course,
the effect is small at lower energies which, however, increases
gradually with energy to yield over 20\% correction.

The parameter $\delta$ (Fig.\,5) being a combination of parameters $\xi$
and $\zeta$ exhibits a rather mixed behavior.
For $2p\thalf$ photoionization (Fig.\,5(a)) a maximum
of 40\% coupling effect is seen at around 300 eV that 
monotonically diminishes with increasing energy to
eventually produce about 20\% effect over the high energy range. For
2p$\ohalf$
(Fig.\,5(b)) the effect of the coupling, that is weak at low energies,
rises steadily to reach a value of about 20\% at the heighest
energy considered.    

These results clearly demonstrate that,
as in the case of cross sections and angular distributions, the 
effect of interchannel coupling on the photoelectron
spin polarization is considerable. However, due to the 
sensitivity of the spin polarization to relative phase-shifts
the results exhibit behavior 
that is qualitatively different from that of the cross section 
results. While for the cross section,
the high energy interchannel coupling effect 
generically increases going from
lower to higher photon energies, for most of the spin polarization
parameters a strong coupling contribution
appears already at low energies. As a consequence, a substantial
coupling correction exists for these parameters 
over a very broad spectral range.  Furthermore, coupling with 1s channels, 
which influences the cross section only over a narrow range, extends its 
influence over a much larger energy range for the 2p$_j$ spin polarization 
parameters owing to the much slower drop-off of the phase shift
corrections 
induced by interchannel coupling. 
Finally, we note that although we have illustrated the effect with
an example of valence photoionization of Ne, the same mechanism
of configuration interactions in the 
{\em continuum} (interchannel coupling) must be operative 
for the spin-polarization of photoejected electrons from any arbitrary
atom or atomic ion and from any subshell. 

\section{\label{con}CONCLUSION}

It is shown in this paper that the photoelectron
spin polarization of an atom is strongly
influenced by the electron correlation {\em via} interchannel 
coupling over the entire spectral range. Unlike to the cross section,
where only the coupling-induced alteration of the magnitudes of matrix
elements is 
responsible for the behavior, modification of the 
phase-shifts play an important role determining the effect on the 
spin polarization. 
For the cross section and the angular distribution the importance
of interchannel coupling for energetic photoemission has been 
demonstrated previously\cite{dias97}.
With the current result we conclude, therefore, that in order
to acquire a {\em complete}
knowledge of photoionization dynamics unambiguously over the 
range from VUV all the way to hard x-rays, theoretical study   
including the interchannel coupling is absolutely required.    

\begin{acknowledgments}
This work is partly supported by NSF, NASA and DST-India. The authors are grateful to
Walter Johnson, University of Notre Dame, for the use of his RRPA code.
HSC acknowledges the computer time from the Department of Physics and
Astronomy, GSU, Atlanta.  
\end{acknowledgments}


\begin{thebibliography}{99}
\bibitem{dias99} E.W.B. Dias, H.S. Chakraborty, P.C. Deshmukh,
                 and S.T. Manson, 
                 J. Phys.\ B {\bf 32}, 3383 (1999).
\bibitem{rado90} V. Radojevic and J.D. Talman, 
                 J. Phys.\ B {\bf 23}, 2241 (1990).
\bibitem{kaem9193} B. K\"{a}mmerling and V. Schmidt, 
                    Phys.\ Rev.\ Lett. {\bf 67}, 1848 (1991);
                    J. Phys.\ B {\bf 26}, 1141 (1993).
\bibitem{snell01} G. Snell, U. Hergenhahn, N. M\"{u}ller, M. Drescher, J.
                  Viefhaus, U. Becker, and U. Heinzmann,
                  Phys.\ Rev.\ A {\bf 63}, 032712 (2001).
\bibitem{fano69} U. Fano,
                 Phys.\ Rev. {\bf 178}, 131 (1969). 
\bibitem{cher83} N.A. Cherepkov, 
                 Adv.\ At.\ Mol.\ Phys. {\bf 19}, 395 (1983).
\bibitem{heinz96} U. Heinzmann and N.A. Cherepkov,
                  in {\em VUV and Soft X-Ray Photoionization}, edited by
                  U. Becker and D.A. Shirley (Plenum, New York, 1996).
\bibitem{beth58} H.A. Bethe and E.E. Salpeter,
                 {\em Quantum Mechanics of One- and Two-Electron Atoms}
                 (Springer-Verlag, Berlin, 1958), Sec.\ 71
\bibitem{coop75} W. Cooper,
                 in {\em Atomic Inner-Shell Processes},
                 edited by B. Crasemann (Academic Press, New York, 1975)
                 Vol.\ 1, p.\ 170.
\bibitem{mans78} S.T. Manson and D. Dill,
                 in {\em Electron Spectroscopy: Theory, Technique and
                 Applications},
                 edited by C.R. Brundle and A.D. Baker (Academic Press,
                 New York,
                 1978) p.\ 186.
\bibitem{berk79} J. Berkowitz,
                 {\em Photoabsorption, Photoionization and Photoelectron
                 Spectroscopy}
                 (Academic Press, New York, 1979) p.\ 61.
\bibitem{star82} A.F. Starace,
                 in {\em Handbuch der Physik},
                 edited by W. Mehlhorn (Springer-Verleg, Berlin. 1982)
                 Vol.\ 31, p.\ 46.
\bibitem{star96} A.F. Starace,
                 in {\em Atomic, Molecular, \& Optical Physics Handbook},
                 edited by G.W.F. Drake (AIP Press, Woodbury, NY, 1996)
                 p.\ 305
\bibitem{dias97} E.W.B. Dias, H.S. Chakraborty, P.C. Deshmukh, S.T.
                 Manson,
                 O. Hemmers, P. Glans, D.L. Hansen, H. Wang, S.B.
                 Whitfield,
                 D.W. Lindle, R. Wehlitz, J.C. Levin, I.A. Sellin, and
                 R.C.C. Perera,
                 Phys.\ Rev.\ Lett. {\bf 78}, 4553 (1997).
\bibitem{hans99} D.L. Hansen, O. Hemmers, H. Wang, D.W. Lindle, P. Focke,
                 I.A. Sellin, C. Heske, H.S. Chakraborty, P.C. Deshmukh,
                 and S.T. Manson,
                 Phys.\ Rev.\ A {\bf 60}, R2641 (1999).
\bibitem{chak01} H.S. Chakraborty, D.L. Hansen, O. Hemmers, P.C. Deshmukh, 
                 P. Focke, I.A. Sellin, C. Heske, D.W. Lindle, and S.T.
                 Manson,
                 Phys.\ Rev.\ A {\bf 63}, 042708 (2001).
\bibitem{john79} W.R. Johnson and C.D. Lin,
                 Phys.\ Rev.\ A {\bf 20}, 964 (1979).
\bibitem{john80} W.R. Johnson, C.D. Lin, K.T. Cheng, and C.M. Lee,
                 Phys.\ Scr. {\bf 21}, 409 (1980). 
\bibitem{huan79} K.-N. Huang, W.R. Johnson, and K.T. Cheng,
                 Phys.\ Rev.\ Lett. {\bf 43}, 1658 (1979).
\bibitem{huan80} Keh-Ning Huang,
                 Phys.\ Rev.\ A {\bf 22}, 223 (1980).
\bibitem{fano86} U. Fano and A.R.P. Rau,
                 in {\em Atomic Collisions and Spectra}
                 (Academic Press, Orlando, 1986) p.\ 66.
\bibitem{comment} From this very simple argument it is easy to show that
                  the full
                  partial cross section for any $nl(l>0)$ 
                  photoionization will always behave as $E^{-9/2}$
                  in the limit $E\rightarrow \infty$ owing to the
                  interaction 
                  with nearby $nl(l=0)$ channels. This corresponds to a
                  part of
                  the result obtained in Ref.\,(\cite{amus00})   
\bibitem{amus00} M.Ya. Amusia, N.B. Avdonina, E.G. Drukarev, S.T. Manson,
                 and R.H. Pratt,
                 Phys.\ Rev.\ Lett. {\bf 85}, 4703 (2000).
\end{thebibliography}
\end{document}